# Atomic origin of high temperature electron trapping in MOS devices

Xiao Shen[1,*] Sarit Dhar[2], and Sokrates T. Pantelides[1,3,4]

[1] Department of Physics and Astronomy, Vanderbilt University, Nashville, TN 37235, USA

[2] Department of Physics, Auburn University, Auburn, AL 36849, USA

[3] Department of Electrical Engineering and Computer Science, Vanderbilt University, Nashville, TN 37235, USA

[4] Materials Science and Technology Division, Oak Ridge National Laboratory, Oak Ridge, TN 37831, USA

ABSTRACT

MOSFETs based on wide band-gap semiconductors are suitable for operations at high temperature, at which additional atomic-scale processes that are benign at lower temperatures can get activated which results in device degradation. Recently significant enhancement of electron trapping was observed under positive bias in SiC MOSFETs at temperatures higher than 150°C. Here we report first-principle calculations showing that the enhanced electron trapping is associated with thermally activated capturing of a second electron by an oxygen vacancy in $SiO_2$, by which the vacancy transforms into a new structure that comprises one Si dangling bond and a bond between a five-fold and a four-fold Si atoms. The results suggest a key role of oxygen vacancies and their structural reconfigurations in the reliability of high-temperature MOS devices.

---

[*] xiao.shen@vanderbilt.edu

Power electronics are key elements in a wide range of technologies such as smart electric grid, [1] renewable energy, [2] and electric vehicles [3]. Power MOSFETs based on wide band-gap semiconductors such as SiC [4] are important types of power electronics that are gaining momentum in both research and commercial arenas. One of the benefits of using wide band-gap semiconductors in power MOS devices is that the devices can in principle operate at higher temperature than their Si-based counterparts, reducing complex and expensive cooling requirements. However, at high temperatures, gate oxide degradation processes that do not happen at lower temperatures can be activated. Therefore, exploring and understanding such high temperature processes are important for establishing and improving the reliability of these devices.

One of the major reliability issues of MOSFETs is the bias-temperature instability (BTI) [5,6]. BTI in Si-MOSFETs has been studied extensively and its atomic origin has been identified as the release of hydrogen from passivated interfacial Si dangling bonds [6]. BTI in SiC-based MOSFETs has been studied by several groups [7-10] and oxygen vacancy was found to play a pivotal role in hole trapping in negative BTI (NBTI) [10]. Recently, Lelis and coworkers reported studies of positive BTI (PBTI) in 4H-SiC MOSFETs and showed that electron trapping is significantly enhanced at temperatures greater than 150 °C [11, 12], which poses great concern to the high temperature reliability of SiC MOS devices. The atomistic origin of the enhanced electron trapping, however, is not yet understood.

In this letter, we report results from first-principle calculations and show that oxygen vacancies in $SiO_2$ are responsible for the observed enhanced electron trapping at high temperature. We show that an oxygen vacancy can capture a second electron through a thermally activated process and transform into a new structure, which comprises one Si dangling bond and a bond between a five-fold and a four-fold Si atoms. This process requires both electron injection and high temperatures, which are satisfied during PBTI. The theoretical results illustrate the important role of oxygen vacancies and their structural reconfigurations in the high-temperature PBTI of MOS devices.

Structures, energies, and electronic properties of oxygen vacancies in $SiO_2$ were investigated using a 72-atom unit of α-quartz. The calculations are based on hybrid density functional theory with the Heyd–Scuseria–Ernzerhof 06 (HSE06) [13] version of the exchange-correlation functional. We use 35% of Hartree-Fork exchange that reproduces the 8.9 eV band gap of $SiO_2$ [14]. We use projector augmented wave potentials (PAW) [15] and plane-wave basis as implemented in the VASP code [16]. The energy cutoff of plane-wave basis is 400 eV. The ionic relaxations are converged to $10^{-3}$ eV for the total energy difference between two steps. A single k-point at (¼, ¼, ¼) is used for Brillouin zone sampling. For the formation energies of charged defect states, the image-charge correction scheme in Ref. 17 was applied. The energy barriers are calculated using the nudged elastic band method [18].

It has been known that oxygen vacancies are prone to form in the silica during thermal oxidation of silicon as the result of thermodynamic fluctuations. In Figure 1 we show the two known structures of an oxygen vacancy $V_O$ [19-22] along with the new structure that accounts for the experimental data. Figure 1a shows the dimer structure of $V_O$ that is formed when one O atom (marked by solid arrow) is removed. Figure 1b shows the puckered structure in which one of the two Si atoms that originally bond to the missing O atom puckers back and bonds with an O atom, forming a 3-fold coordinated O (solid arrow), while the other Si atom retains a Si-dangling bond (dashed arrow). This structure plays a key role in hole-trapping-related phenomena in $SiO_2$ [10, 20-22]. Figure 1c is the new configuration of oxygen vacancy. It also has one Si atom with a dangling bond (dashed arrow), but instead of having a 3-fold O, it contains a Si-Si bond between a five-fold coordinated Si and a four-fold coordinated Si. This structure bears similarity to one of proposed models for a hydrogenated oxygen vacancy [SiH-SiSi(5)] [23], but in the present case, neither of the Si atoms in the Si-Si bond is initially bonded to the missing O atom. We name this structure $Si^5$-$Si^4$ and show that the reconfiguration of $V_O$ into this structure could be responsible for the PBTI results at high temperature.

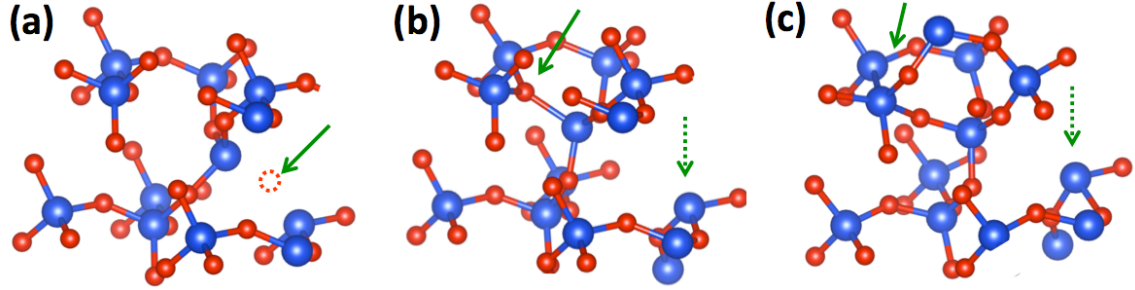

Figure 1. Three possible structures of the oxygen vacancy $V_O$. (a) The dimer structure, with solid green arrow marking the missing O atom. (b) The puckered structure, with solid green arrow marking the 3-fold O atom and dashed green arrow marking the Si dangling bond. (c) The new $Si^5$-$Si^4$ structure, with solid green arrow marking the bond between a 5-fold and a 4-fold Si and dashed green arrow marking the Si dangling bond.

Figure 2 shows the formation energies of $V_O$ at various charge states in the three possible structures shown in Fig. 1. When SiC MOS devices are at resting state, the Fermi level is near the middle of the $SiO_2$ band gap, therefore, the charge neutral dimer configuration is most stable. During positive bias stress, the quasi-Fermi level in $SiO_2$ is close to its conduction band, as free electrons are injected into the oxide from the inversion layer in n-channel 4H-SiC MOSFETs. Under this condition, the $Si^5$-$Si^4$ structure is the thermodynamic ground state. However, the trapping of electrons at oxygen vacancies likely happens in sequel, so that $V_O^{2-}$ ($Si^5$-$Si^4$) can only be formed when a singly negatively charged $V_O^-$ (dimer) captures one more electron from the $SiO_2$ conduction band and undergoes a structure transformation. Figure 3 shows the configuration diagram of such process, from which we obtain an activation energy of ~2 eV. In amorphous $SiO_2$, the reaction barriers at different sites can have a spread in energy of about 0.7 eV [24]. Therefore we estimate the activation energy to spread between 1.6 to 2.3 eV in amorphous $SiO_2$. The lower end of the estimated activation energy (1.6 eV) corresponds to an activation temperature around 150 °C. As a result, the formation of $V_O^{2-}$ ($Si^5$-$Si^4$) should begin to occurs at temperature above 150 °C, which is consistent with the observed enhancement of electron trapping.

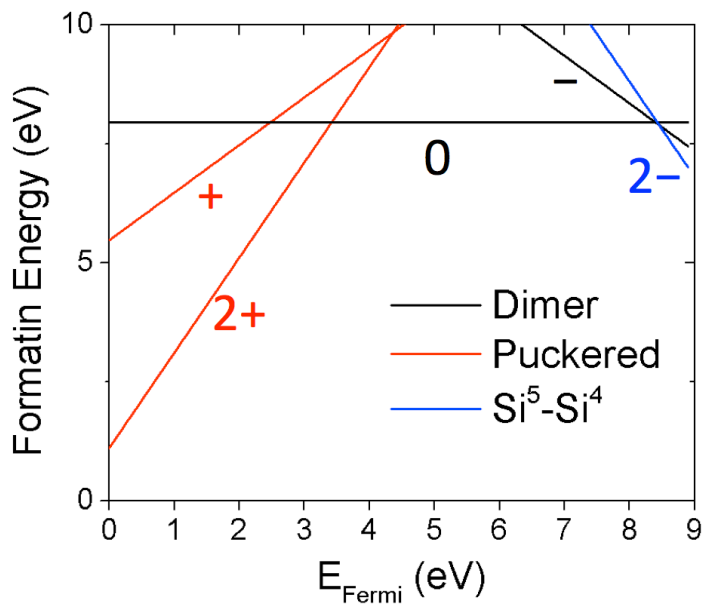

Figure 2. Formation energy of oxygen vacancy in SiO$_2$ with the charge states labeled.

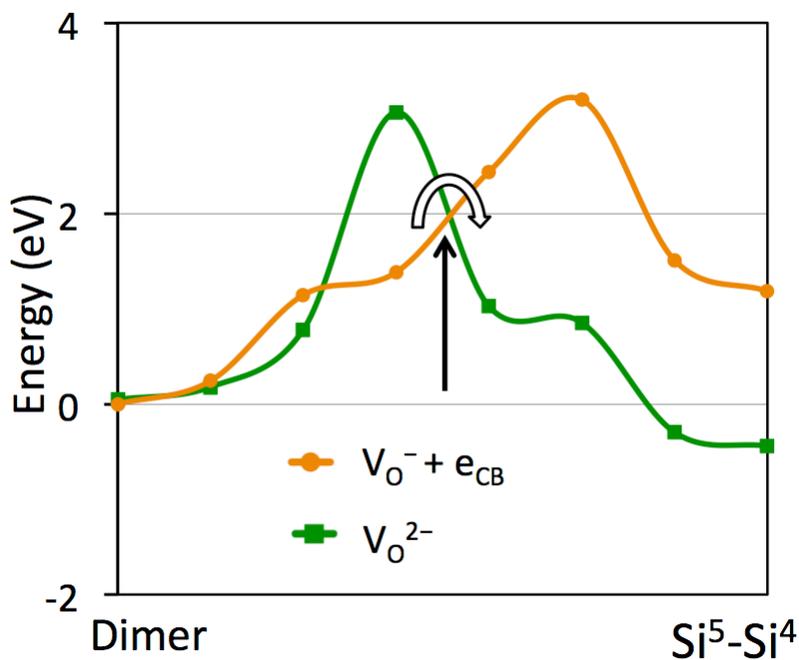

Figure 3. Configuration diagram of oxygen vacancy transforming from dimer structure to the Si$^5$-Si$^4$ structure. The orange curve represents the energy of V$_O^-$ plus one electron at the bottom of the SiO$_2$ conduction band (CB). The green curve represents the energy of V$_O^{2-}$. The white arrow shows the transition from $V_O^-$ (dimer) + $e_{CB}$ to $V_O^{2-}$ ($Si^5$-$Si^4$). The black arrow shows the activation energy.

Once an oxygen vacancy traps a second electron and becomes $V_O^{2-}$ ($Si^5$-$Si^4$), it requires 2.4 eV of energy to emit an electron back to the conduction band and revert to $V_O^-$ (dimer). (For amorphous SiO2, we estimate that the lower end of the activation energy is 2.0 eV.) Therefore, the double-negatively charged oxygen vacancy $V_O^{2-}$ ($Si^5$-$Si^4$) is very stable against electron emission. Nevertheless, at sufficient high temperature (> 250°C), this process can still happen, converting $V_O^{2-}$ ($Si^5$-$Si^4$) to $V_O^-$ (dimer), which can then emit another electron and become $V_O^0$ (dimer). This suggests that the trapped electrons can be annealed at high temperatures. Furthermore, $V_O^{2-}$ ($Si^5$-$Si^4$) can be easily neutralized when holes are available to tunnel from SiC into $SiO_2$. This means that the electron trapping during PBTI can be recovered by applying a negative bias to the MOSFET, which can create accumulation of holes at the interface and injects them into the oxide. Compared with Si, the SiC MOS devices undergo higher oxidation temperature and are likely to have more oxygen vacancies. As a result, the SiC MOS devices are more prone to the degradation mechanism described above.

Now we discuss the electronic structure of an oxygen vacancy in the $V_O^{2-}$ ($Si^5$-$Si^4$) state. In such configuration, the $V_O$ has two energy levels within the band gap of $SiO_2$, as shown in Figure 4. The Si dangling bond has an energy level at 3.8 eV above $SiO_2$ VBM, which is 0.9 eV above the VBM of 4H-SiC (the valence band offset of SiC and $SiO_2$ is 2.9 eV [25]). Therefore, those oxygen vacancies that are located close to the interface should be accessible by capacitance-voltage (CV) measurements as they can exchange charge with the SiC substrate by tunneling. (Although the defect level of 0.9 eV calculated in α-quartz is quite deep to be accessed by CV measurement, in amorphous $SiO_2$ there exists a distribution of energy levels and some of them are shallow enough to be probed by CV through charge exchange. For example, the Si dangling bond state of the four-fold puckered $V_O$ in amorphous $SiO_2$ has a level at 0.6 eV above the SiC VBM [24]). On the other hand, the energy level associated with the Si-Si bond is 2.1 eV above the $SiO_2$ VBM, which is 0.8 eV below the VBM of 4H-SiC and thus inaccessible by the CV measurements.

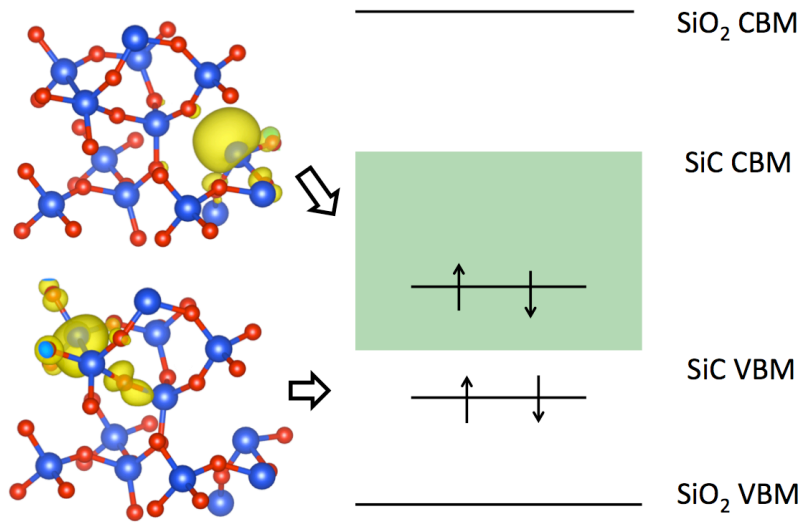

Figure 4. Wavefunctions (shown as yellow isosurfaces) and associated energy levels of an oxygen vacancy in the $V_O^{2-}$ ($Si^5$-$Si^4$) state. The range of 4H-SiC band gap is marked in green.

In summary, we have shown that the enhanced electron trapping in 4H-SiC MOS devices under positive bias at temperatures greater than 150 °C originates from a thermally activated structural transformation of oxygen vacancies upon capturing a second electron. This process is intrinsic to $SiO_2$ and therefore should also be relevant in MOS devices other than SiC MOSFETs. In order to achieve good reliability in high-temperature MOS device, it would be essential to reduce the amount of oxygen vacancies in the oxide. Low temperature anneal in oxygen-rich ambient may reduce the oxygen vacancy concentration and improve the high temperature reliability.

**Acknowledgement**

The work at VU was supported in part by NSF GOALI grant DMR-0907385 and by the McMinn Endowment at Vanderbilt University. Computational resources are provided by the National Science Foundation through XSEDE resources under grant number TG-DMR100022 and by NERSC, supported by DOE under Contract No. DE-AC02-05CH11231. The work at AU was supported by the US Army Research Laboratory (W911NF-07-2-0046, Program Manager: Dr. Aivars Lelis). We thank B. R. Tuttle for helpful discussions.